\documentclass{article}

\usepackage{arxiv}

\usepackage[utf8]{inputenc}
\usepackage[T1]{fontenc}
\usepackage{hyperref}
\usepackage{amsfonts}
\usepackage{amsmath, amssymb, amsthm}
\usepackage{microtype}
\usepackage{graphicx}
\usepackage{bm}
\usepackage{authblk}
\usepackage{enumitem}
\usepackage{xcolor}

\title{Optimal multi-asset trading with linear costs: \\ a mean-field approach}

\author[1,3]{{\large\textbf{Matt Emschwiller}}}
\author[2,3]{{\large\textbf{Benjamin Petit}}}
\author[4,5]{{\large\textbf{Jean-Philippe Bouchaud}}}

\affil[1]{\textit{Operations Research Center}, Massachusetts Institute of Technology, Cambridge, MA}
\affil[2]{\textit{Institute for Computational and Mathematical Engineering}, Stanford University, Stanford, CA}
\affil[3]{\textit{Ecole polytechnique}, Route de Saclay, 91128 Palaiseau Cedex, France}
\affil[4]{\textit{Capital Fund Management}, 23 Rue de l'Universit\'e, 75007 Paris, France}
\affil[5]{\textit{CFM-Imperial Institute of Quantitative Finance}, Dept of Mathematics, Imperial College, London SW7 2AZ, UK \vspace{-0.25in}}

\begin{document}

\maketitle

\begin{abstract}
Optimal multi-asset trading with Markovian predictors is well understood in the case of quadratic transaction costs, but remains intractable when these costs are linear in the quantity traded. We present a mean-field approach that reduces the multi-asset problem to a single-asset problem, with an effective predictor that includes a risk averse component. We obtain a simple approximate solution in the case of Ornstein-Uhlenbeck predictors and maximum position constraints. The optimal strategy is of the ``bang-bang'' type similar to that obtained in \cite{Lataillade}. When the risk aversion parameter is small, we find that the trading threshold is an affine function of the instantaneous global position, with a slope coefficient that we compute exactly. We relate the risk aversion parameter to the desired target risk and provide numerical simulations that support our analytical results.
\end{abstract}

\section{Introduction}

Classical financial theory asserts that future price moves are unpredictable. In fact, all sorts of anomalies have been found that contradict the efficient market hypothesis: prices display some degree of predictability, which can be exploited by asset managers through statistical arbitrage strategies.

However, transaction costs (fees, spreads and market impact) make these strategies unprofitable when the underlying predictive signals are not exploited properly. Optimal trading in the presence of costs has become the focus point of the asset management industry, accompanied by a growing academic literature \cite{Almgren,Cartea,Gueant,TQP,Lintilhac,Lehalle,Belak}. One famous example is the G\^arleanu-Pedersen solution for optimal trading in the presence of quadratic costs (i.e. costs that grow as the square of the trading speed). In that case, the solution essentially amounts to resizing and slowing down the predictor, using an exponential moving average with an appropriate friction \cite{Garleanu}. The problem is much less analytically tractable in the case of linear costs (i.e. trading fees or bid-ask spread). Trading in that case becomes discontinuous: when the signal is too small, it is better not to trade at all, see e.g. \cite{Davis, Shreve, Martin, Lataillade}. More complicated cases, with both linear and quadratic costs, or more general cost functions, have been considered as well, see e.g. \cite{JMK1, JMK2, Rej}.

Unfortunately, most of the results in the linear cost case concern the single asset problem. Determining the shape of the no-trade region in the multi-asset case is a quite fascinating problem, for which no analytic solution exists in the general case, even in the limit for small transaction costs treated by Possamai, Soner and Touzi \cite{Possamai}. The aim of the present paper is to extend the formalism of de Lataillade et al. \cite{Lataillade} to treat the multi-asset problem with linear transaction costs, a maximum position constraint on each asset and a quadratic portfolio risk penalty. We propose a ``mean-field'' approach to the problem (in a sense made clear later) that we conjecture becomes exact in the limit of large portfolios, provided the number of common risk factors remains finite. Such ``mean-field'' approaches have been successfully applied to other problems, such as the optimal \textit{liquidation} problem \cite{Cardaliaguet,Firoozi} in a multi-agent setting. In our case, the problem boils down to the single-asset problem solved in \cite{Lataillade}, with a modified predictor that includes a (mean-field) risk contribution. The statistics of the modified predictor depends on the solution of the problem and must be self-consistently determined. 

We provide a solution of the mean-field problem in the case of independent Ornstein-Uhlenbeck predictors for each asset and a single common risk factor. In this case, the risk-aware predictor is the sum of two independent Ornstein-Uhlenbeck processes, for which the boundary of the no-trade zone cannot be obtained exactly in full generality. Approximate solutions are found in the empirically relevant limit where the risk aversion term is small compared to the prediction component. We find that the boundary of the no-trade region is an affine function of the risk, with a slope coefficient that we compute. We also provide numerical simulations that support our analytical calculations. Several extensions of our work are suggested in the conclusion.  

\section{Background: single-asset optimal trading}

\subsection{Formulation of the problem} \label{formulation}

In \cite{Lataillade}, the authors consider the following one-asset optimal trading problem in discrete time. An investor must determine his/her signed position $\pi_t$ at (integer) time $t$, when he/she receives a signal $p_{t}$ that predicts the next price change $r_{t}=\text{price}_{t+1}-\text{price}_{t}$, with the following constraints:
\begin{itemize}
    \item The signal is such that $\mathbb{E}\left[r_{t}|p_{t}\right]=p_{t}$ (the true return is centered around the predictive signal), and follows a discrete Ornstein-Uhlenbeck (aka AR(1)) process, i.e. $p_{t+1}-p_{t}=-\epsilon p_{t}+\psi\xi_{t}$, where $\xi_t$ are iid, $N(0,1)$ random variables.\footnote{More general Markovian predictors were considered in \cite{Lataillade}, but we will restrict the current paper to AR processes.} 
    \item The risk control is a cap on the absolute size of the position: $\forall t,\,|\pi_{t}|\leq M$.
    \item Each transaction of size $Q=|\pi_{t+1}-\pi_{t}|$ generates some linear cost $=\Gamma Q$.
\end{itemize}
The investor wants to maximize his/her long term expected gain, by choosing optimally the sequence of positions $\pi_t$ over a period $[0,T]$, with $T\rightarrow+\infty$. In other words, we are interested in the ergodic limit of the problem.

\subsection{The two-threshold solution}

At first glance, a reasonable approach could be to take a position $\pm M$ whenever the predictive signal $|p_{t}|$ exceeds the trading cost $\Gamma$. This strategy generates a positive gain, but there is no reason for it to be optimal. In fact, this strategy does not use the auto-correlation time (the length of its memory) of the signal $p_{t}$, nor its persistence. It is paramount to observe that a decision taken now will influence future trading and future transaction costs. By trading at a loss at some time $t$ (where the transaction cost $\Gamma$ would be higher than the expected reward $p_t$), the trader can count on the fact that the predictive signal $p$ will stay in a neighborhood of $p_t$ for some time, leading to a total future profit $p_t+p_{t+1}+...$ which hopefully will be greater than the initial cost $\Gamma$. Specifically, the agent tries to find a deterministic policy that maximizes the following value function, defined recursively:
\begin{align*}
    V_t(\pi, p) = \max_{|\pi_{t+1}| \leq M} \left( p \cdot \pi_{t+1} - \Gamma|\pi_{t+1} - \pi| + \int P\left( p_{t+1}=p' | p_t=p \right) V_{t+1}(\pi',p')dp' \right)
\end{align*}
In the infinite horizon limit ($T \rightarrow +\infty$), De Lataillade et al. \cite{Lataillade} solved this Hamilton-Jacobi-Bellman equation and proved that the optimal discrete time trading strategy $\pi^*$ is: 
\begin{equation}
    \pi_{t}^{*} = \begin{cases}
    \pi_{t-1}^{*} & \text{if } |p_{t}|<q^* \\
    M & \ \text{if } p_{t}\geq q^*\\
    -M & \ \text{if } p_{t}\leq -q^*.
    \end{cases}	
\end{equation}
The optimal threshold $q^{*}$ satisfies a certain equation which can be interpreted in the following manner: the expected profit obtained from switching from position $-M$ to position $+M$ when the predictor is just below $q^*$ and remains in $[-q^{*}, q^{*}]$  must be equal to $2 \Gamma$ times the probability to reach $-q^{*}$ without ever touching $q^{*}$, see \cite{Lataillade} and section \ref{sec:optimal_thresholds} for details. In fact, our solution below, Eqs. \eqref{eq_q}, \eqref{eq_q_cases}, recovers the solution of Lataillade et al. when risk aversion is zero ($\theta=0$). 

\begin{figure}[htbp]
\centering
\includegraphics[width=0.5\textwidth]{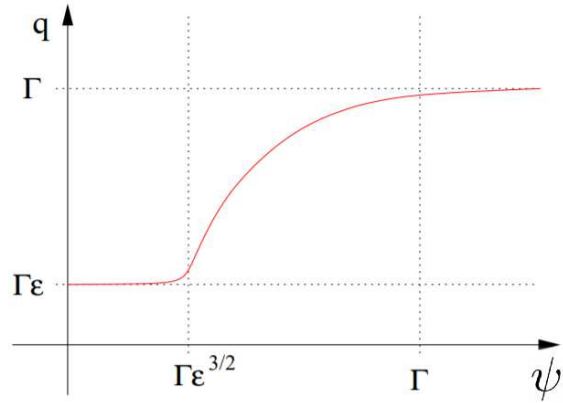}
\caption{Threshold $q^*$ as a function of the predictability $\psi$ for a discrete Ornstein-Uhlenbeck predictor (from \cite{Lataillade}).}
\label{Single_asset_solution_regimes}
\end{figure}

When the predictor follows an AR(1) process, one can derive an exact formula for $q^*$, which reveals three regimes -- see Fig. \ref{Single_asset_solution_regimes}:
\begin{enumerate}[label=(\alph*)]
\item Weak predictabilities $\left[\psi \ll \Gamma \epsilon^{3/2}\right]$: $q^* \approx \Gamma \epsilon$ (note that $\psi \epsilon^{-3/2}$ is the order of magnitude of the total predicted gain).
\item Intermediate predictabilities $\left[\Gamma \epsilon^{3/2} \ll \psi \ll \Gamma\right]$:
\begin{equation}\label{lataillade_threshold}
    q^{*}=\left(\frac{3}{2}\Gamma\psi^{2}\right)^{\frac{1}{3}}.
\end{equation}
This corresponds to the continuous-time limit result, which is also the most interesting in practice. The $\Gamma^{1/3}$ dependence of the threshold first appeared in \cite{Shreve}, see also \cite{Martin,JMK2}. 
\item Strong predictabilities $\left[\psi \gtrsim \Gamma\right]$: $q^{*} \approx \Gamma$. This is for instance the regime corresponding to a white noise predictor ($\epsilon \approx 1$), for which the intermediate regime disappears. The absence of correlation between time steps means that the predictor has to beat the linear cost for a trade to take place. 
\end{enumerate}

\section{The multi-asset discrete Ornstein-Uhlenbeck problem}

In this section, we extend the framework described in \cite{Lataillade} to a multi-asset setting.

\subsection{The set-up}

We now consider trading a universe of $N$ financial assets, $i=1,\dots,N$, where $N$ is a large number (typically, a few hundreds/thousands). We assume that the returns of these 
assets are described by a covariance matrix $C_{ij}$. Without loss of generality, we assume that the volatility of these assets are all equal to unity (we can always rescale the positions $\pi^i$ for this to be the case).

We also assume that the predictors $p^i_t$ for each asset only concern the idiosyncratic part of its evolution, and are thus mutually independent. Their dynamics are given by independent discrete Ornstein-Uhlenbeck processes:
\begin{equation} \label{p_dynamics}
    p_{t+1}^{i}-p_{t}^{i}=-\epsilon_i p_{t}^{i}+\psi_i \xi_{t}^{i},
\end{equation}
with predictabilities $\psi_i$ and correlation times $1/\epsilon_i$ possibly all different. Risk control is enforced through two types of constraints: a maximum position per asset 
$|\pi^i| \leq M_i$ that limits the idiosyncratic risk, and the standard portfolio risk penalty $\mathcal{R}= \sum_{ij} \pi^i C_{ij} \pi^j /N$ (note that this is a risk {\it per asset}). Our investor wants to maximize his/her long term gain, factoring in linear costs and risks, i.e. (with self-explanatory notations):
\begin{equation}
    \mathcal{G}_{[0,T]}=\sum_{t=0}^{T-1} \left[\vec p_t \cdot \vec \pi_{t}-\sum_i \Gamma_i |\pi^i_{t+1}-\pi^i_t| -\lambda \mathcal{R}_t\right],
\end{equation}
under the constraint $|\pi^i| \leq M_i$, $\forall i$, and where $\lambda$ is a risk aversion parameter. The corresponding HJB equation is obtained by introducing a value function $\mathcal{V}_t(\vec \pi, \vec p)$ that obeys the following recursion equation:
\begin{multline}
    \mathcal{V}_{t}(\vec \pi,\vec p)=\max_{|\pi^i_{t+1}|\leq M_i}\left[\vphantom{\int_1^2}\vec p\cdot\vec \pi_{t+1}-\sum_i \Gamma_i |\pi^i_{t+1}-\pi^i|-\frac{\lambda}{N} \vec \pi_{t+1}^\top \mathbf{C} \vec \pi_{t+1}\right. \\ \left. +\int P(\vec p_{t+1}=\vec{p'}|\vec p_{t}=\vec p)\mathcal{V}_{t+1}(\vec \pi_{t+1},\vec{p'})d\vec{p'}\vphantom{\int_1^2} \right],
\end{multline}
with -- say -- the constraint that the portfolio is empty when $t=T$, i.e. $\vec \pi_T = 0$. Since we will be interested in the limit $T \to \infty$, the precise boundary condition is in fact irrelevant in what follows. Note that general problem boils down to the one considered in \cite{Lataillade} when $N=1$ and $\lambda=0$. 

\subsection{The mean-field limit}

In order to make progress, we henceforth assume that the covariance matrix derives from a simple one-factor model, such that $C_{ii}=\sigma_i^2 + \beta_i^2$ and $C_{ij}=\beta_i \beta_j$ ($i \neq j$), where $\beta$ are the usual ``beta'' to the market. (We conjecture that our method applies to more complicated covariance matrices with a hierarchical block structure, provided the number of blocks remains small compared to $N$). Note that as soon as the portfolio acquires a net position $(\sum_{i}{\beta_i\pi^i}=O(N))$, the risk per asset becomes of order $N$, and the risk penalty 
$\lambda \mathcal{R}$ is of the same order in $N$ as the gain and the cost terms in the value function when $\lambda = O(1)$. This is the regime we will study throughout this paper.

For this particular choice of risk model, one can rewrite the risk per asset as:
\begin{equation}
    \mathcal{R}_t = \frac{1}{N} \left[ \sum_{i=1}^{N} C_{ii} (\pi_{t}^{i})^2 + \sum_{\substack{i,j=1 \\ i \neq j}}^{N} \beta_i \beta_j \pi_{t}^{i} \pi_{t}^{j} \right],
\end{equation}
which we rewrite as
\begin{equation}\label{eq_risk}
    \mathcal{R}_t = \frac{1}{N} \sum_{i=1}^{N} C_{ii} (\pi_{t}^{i})^2 + \frac{1}{\sqrt{N}} \sum_{i=1}^{N} \beta_i \pi_{t}^{i} R_t^i,
\end{equation}
with 
\begin{equation}
R^i_{t}= \frac{1}{\sqrt{N}}\sum_{j \neq i} \beta_j \pi_{t}^{j}.
\end{equation}
Note that the market risk (modelled by the second term in Eq. \eqref{eq_risk}) is, when $N$ is large, much larger than the idiosyncratic contribution (i.e. the first term), which we will neglect henceforth.

The central idea of the mean-field in statistical mechanics is that for large $N$, $R_t^i$ can be treated as an independent random variable whose dynamics is self-consistently determined, assuming that the $N$-asset problem can be decomposed into $N$ (quasi)-independent one-asset problems, only coupled through the value of the ``mean-field'' $R_t:=\sum_{j} \beta_j \pi_{t}^{j}/\sqrt{N}$, which represents the total beta-weighted net position of the portfolio:
\begin{equation}
    V_t(\vec \pi_{t},\vec p_t, R) = \sum_{i=1}^N {V}^i_{t}(\pi_t^i,p_t^i,R),
\end{equation}
with modified, ``risk-aware'' predictors and value functions $V^i_t(\pi,p,R)$ that obey the following recursion relations:
\begin{multline}\label{opt_prob}
    {V}^i_{t}(\pi,p,R)=\max_{|\pi_{t+1}|\leq M_i}\left[\vphantom{\int_1^2} (p - \theta \mathbb{E}[R_{t+1}|R] )\pi_{t+1}-\Gamma_i |\pi_{t+1}-\pi| \right. \\ \left. +\int P(p_{t+1}=p',R_{t+1}=R'|p_{t}=p,R_t=R){V}^i_{t+1}(\pi_{t+1},p',R')dp' dR' \vphantom{\int_1^2} \right],
\end{multline}
where $\theta:=\lambda \beta/\sqrt{N}$. The subtle point is that the statistics of $R_t$ has to be self-consistently determined by the solution to the one-asset problems, each of which is solvable -- in principle -- using the formalism of de Lataillade et al. \cite{Lataillade}. However, this task is still daunting because the modified predictor $p - \theta \mathbb{E}[R_{t+1}|R]$ cannot be characterized in general.

Our goal is to solve the problem in the limit $\theta \to 0$. In fact, we will see in section \ref{sec:fixing_lambda} below that in order to determine the average realized risk, the risk aversion parameter $\lambda$ must indeed be of order $1$. This is the regime in which the net position $R_t$ also remains $O(1)$, which is the interesting limit for financial applications in the context of portfolios with offsetting long/short positions. In this regime, $\theta$ indeed becomes very small and the calculations below are fully justified. In fact, we will show that our approach is valid as long as $\lambda=o(\sqrt{N})$, see Eq. \eqref{eq_cond} below.

In the next section, we first show that, in the continuous time limit and when $N\rightarrow +\infty$, $R$ is itself an Ornstein-Uhlenbeck process with computable parameters. 

\section{Asymptotic, continuous time dynamics of the mean-field} \label{sec:mean_field_dynamics}

Our strategy is to derive the dynamics of $R$ in the limit $N\rightarrow +\infty$, when each single asset problem formally boils down to the problem solved in \cite{Lataillade} when $\theta \to 0$. For simplicity, we will restrict henceforth to the {\it continuous time regime} (corresponding to the intermediate predictibility regime (b) in our classification above). In this regime, the  time step ``1'' becomes an infinitesimal quantity $dt$ and the predictors can be treated as Ornstein-Uhlenbeck processes, i.e  
\begin{equation} \label{p_dynamics2}
    dp_{t}^{i}=-\epsilon_i p_{t}^{i}dt+ \psi_i dW_{t}^{i}
\end{equation}
where $\left(W^{1},...,W^{N}\right)$ are independent Wiener processes. For each asset, the single-asset optimal strategy from \cite{Lataillade} is characterized by an optimal threshold $q^*_i > 0$, and is such that $\pi_t^i \in \{\pm M_i\}$ for all $i,t$ (possibly after a short transient that we neglect in the following). By symmetry of the predictors, we have:
\begin{equation}
    \mathbb{P}(\pi_{t}^{i}=M_i)=\mathbb{P}(\pi_{t}^{i}=-M_i)=\frac12.
\end{equation}
The positions are thus Bernoulli random variables. For a given $t$, the $(\pi_{t}^{i})_{i=1...N}$ are independent, as $\pi_{t}^{i}$ is solely a function of $(p_{s}^{i})_{s\in[0,t]}$ (which are assumed to be independent). Using the central limit theorem, we obtain that for all $t$, $R_{t}=\sum_i \beta_i \pi_t^i/\sqrt{N}$ is asymptotically ($N\rightarrow +\infty$) Gaussian. In fact, in what follows, we will prove that $R_{t}$ asymptotically behaves as an Ornstein-Uhlenbeck process when $N\rightarrow+\infty$.

Let us now write the infinitesimal variation of $R_{t}$ as
\begin{equation}
    R_{t+dt}-R_{t}=\frac{2}{\sqrt{N}}\sum_{i=1}^{N} \beta_i M_i \xi_{i},	
\end{equation}
where $\xi_{i}=1$ if position $i$ switches from $-M_i$ to $+M_i$ between $t$ and $t+dt$, $-1$ if position $i$ switches from $+M_i$ to $-M_i$, and $0$ otherwise.
The distribution of $\xi_i$ conditioned to a certain value of $R_t$ is given by:
\begin{equation}
    \begin{cases}
    \mathbb{P}(\xi_{i}=-1|R_{t})=\mathbb{P}(p^i_{t+dt}<-q^*_i,\  p^i_t > -q^*_i|\pi^i_{t}=M_i) \,\, \mathbb{P}(\pi^i_{t}=M_i|R_{t}),\\
    \mathbb{P}(\xi_{i}=1|R_{t})=\mathbb{P}(p^i_{t+dt}>q^*_i,\  p^i_t < q^*_i|\pi^i_{t}=-M_i) \,\, \mathbb{P}(\pi^i_{t}=-M_i|R_{t}),
    \end{cases}
\end{equation}
where the first term in the right hand side corresponds to the (stationary) probability that $p_i$ crosses the threshold $q^*_i$ between $t$ and $t+dt$, that we will call $J_i \, dt$ and compute explicitly in the next section. It corresponds to the rate of trading of asset $i$. 

The second term is the conditional probability to find asset $i$ at its maximum (resp. minimum) position $\pm M_i$, for a given value of $R_t$. In Appendix \ref{app:cond_pi_R}, we show the following result, valid for large $N$:
\begin{equation}
\mathbb{P}(\pi^i_{t}= \pm M_i|R_{t}) = \frac12 \left[1 \pm \frac{R_t \beta_i M_i}{\Sigma^2\sqrt{N}}\right],
\end{equation}
where $\Sigma^2:=\sum_j \beta_j^2 M_j^2/N$. We therefore obtain
\begin{equation}\label{moments}
\mathbb{E}[\xi_{i}|R_{t}]=-\frac{R_{t}\beta_i M_i}{\Sigma^2 \sqrt{N}} J_i \, dt + o(dt), \qquad \mathbb{V}\left[\xi_{i}|R_{t}\right]=J_i \, dt + o(dt).
\end{equation}
Hence, since $(\xi_{i})_{i=1..N}$ are independent variables, $R_{t+dt}-R_{t}$ is asymptotically Gaussian, with the following moments: 
\begin{equation}
    \begin{cases}
    \mathbb{E}\left[R_{t+dt}-R_{t}|R_{t}\right]= -2 \bar{J} R_{t}dt\\
    \mathbb{V}\left[R_{t+dt}-R_{t}|R_{t}\right] = 4 \Sigma^2 \bar{J} dt,
    \end{cases}	
\end{equation}
where $dt \to 0$ and
\[
\bar{J}:= \frac{\sum_i \beta_i^2 M_i^2 J_i}{\sum_i \beta_i^2 M_i^2}
\]
is the average rate of trading, weighted by the square of the maximum positions and the beta's.

In the limit $dt \to 0$, the mean-field risk is thus an Ornstein-Uhlenbeck process with known parameters $\Sigma$ and $\bar{J}$:\footnote{Making this informal convergence argument rigorous using tools such as the ones developed in, e.g. \cite{Whitt}, is a interesting direction for future research.}
\begin{equation} \label{R_dynamics}
    dR_{t}=-2 \bar{J} R_{t}dt+2 \Sigma \sqrt{\bar{J}}dW_{t}. 
\end{equation}
From this expression, one can compute the stationary variance of $R_t$, given by
\begin{equation}
    \mathbb{V}\left[R_{t}\right]=\Sigma^{2},
\end{equation}
which is independent of $\bar{J}$ when $\theta \to 0$. Note that the variance of the predictor $p^i$ is given by
\begin{equation}
    \mathbb{V}\left[p^i_{t}\right]=\frac{\psi_i^{2}}{2\epsilon_i}:=p_i^{*2}.
\end{equation}
The order of magnitude of the risk aversion term $\theta R_t$ is therefore $\theta \Sigma$, to be compared to the scale of the predictors, $p_i^* = \psi_i/\sqrt{\epsilon_i}$. We will consider below the limit $\theta \to 0$, which must in fact be understood as 
\begin{equation}\label{eq_cond}
{\lambda} \ll \frac{p_i^*}{\beta_i \Sigma} \sqrt{N},
\end{equation} 
for all $i=1,...,N$, i.e. the portfolio risk contribution remains small compared to the prediction strength. 

In conclusion of this section, we have shown that provided each asset is traded using an independent Ornstein-Uhlenbeck signal with a no-trade region, the resulting mean-field risk 
term is itself an Ornstein-Uhlenbeck process. This conclusion is valid for a larger class of Markovian predictors as well, and for the full self-consistent problem where each asset sees a modified predictor that includes the risk term, since for this problem too the optimal solution has a ``bang-bang'' nature with a no-trade region. Eq. \eqref{moments} will therefore still be valid, but the explicit computation of $\bar{J}$ is simple only when $\theta \to 0$, to which we now turn.

\section{A theory for the trading rate} \label{sec:th_rates}

In this section, we want to compute the trading rate $J_i$ for each asset $i$, assuming that trading is dominated by the predictors $p_i$ (i.e. when $\theta \to 0$). For notational simplicity, we temporarily drop the index $i$. Let us now introduce the following conditional probability 
\begin{equation}
    Q_{+}(p,t) dp :=\mathbb{P}\left(p_{t}\in [p,p+dp]|\pi_{t}= M\right).	
\end{equation}
Since we assume that $p_t$ follows a continuous-time Ornstein-Uhlenbeck process, the stationary density $Q_{+,\text{st.}}(p,t)$ satisfies a modified Fokker-Planck equation (where $\delta$ is a Dirac mass):
\begin{equation}\label{FP}
    0 = \frac{\partial Q_{+,\text{st}}}{\partial t}=\epsilon\frac{\partial(pQ_{+,\text{st}})}{\partial p}+\frac{\psi^{2}}{2}\frac{\partial^{2}Q_{+,\text{st}}}{\partial p^{2}}+J \, \delta(p-q^*),
\end{equation}
with a boundary condition $Q_{+,\text{st.}}(p=-q^*,t)=0$, describing the fact that when $p$ touches the lower threshold $-q^*$, the position flips from $+M$ to $-M$. This occurs with a rate $J$, given by the probability flux at $p=-q^*$:
\begin{equation}\label{transmut}
J:=\frac{\psi^{2}}{2}\frac{\partial Q_{+,\text{st}}}{\partial p}|_{p=-q^*}.
\end{equation}
The last term in Eq. \eqref{FP} corresponds to the inverse process: the change of $-M$ positions into $+M$ positions when the predictor touches $+q^*$, which occurs at the same rate $J$ (by symmetry). 

The solution of this equation leads to
\begin{equation}
    \begin{cases}
    \epsilon pQ_{+,\text{st.}}+\frac{\psi^{2}}{2}\frac{\partial Q_{+,\text{st.}}}{\partial p}=J & \ \mbox{for}\ -q^*<p<q^*\\
    \epsilon pQ_{+,\text{st.}}+\frac{\psi^{2}}{2}\frac{\partial Q_{+,\text{st.}}}{\partial p}=0 & \ \mbox{for}\ p>q^*\\
    Q_{+,\text{st.}}(-q^*)=0,
    \end{cases}
\end{equation}
which must also be such that $\int_{-q^*}^\infty Q_{+,\text{st.}}(p) dp=1$. Introducing $a:=\epsilon\psi^{-2}=p^{*-2}$, we finally obtain:
\begin{equation}
    Q_{+,\text{st.}}(p)=\frac{2J}{\psi^{2}}e^{-ap^{2}}\int_{-q^*}^{\min(p,q^*)}e^{ax^{2}}dx,
\end{equation}
which is indeed such that $J=\psi^2/2 Q_{+,\text{st.}}'(p=-q^*)$. The normalization of $Q_{+,\text{st.}}$ then leads to:
\begin{equation} \label{FullJ}
\frac{1}{J}=\frac{2}{\psi^{2}}\left[\int_{-q^*}^{q^*}e^{-ap^{2}}\left(\int_{-q^*}^{p}e^{ax^{2}}dx\right)dp+\int_{q^*}^{+\infty}e^{-ap^{2}}dp\int_{-q^*}^{q^*}e^{ap^{2}}dp\right].
\end{equation}
We now introduce $\hat{q}:=q^*/p^*$ as the ratio between the width of the no-trade region and the typical scale of the predictor (ignoring the $\sqrt{2}$ constant). Eq. \eqref{FullJ} can be approximately solved in the two regimes $\hat{q} \ll 1$ (small band) and $\hat{q} \gg 1$ (large band), leading to (see Appendix \ref{app:J_expansion} for more details):
\begin{equation}
\begin{cases}
J \approx \frac{\epsilon}{2\sqrt{\pi}} \hat{q}^{-1} & \ \hat{q}\ll 1 \\
J \approx \frac{\epsilon}{\sqrt{\pi}}\hat{q}e^{-\hat{q}^{2}} & \ \hat{q}\gg 1.
\end{cases}
\label{J_approx}
\end{equation}
Note that $q^*\ll p^* \iff J \gg \epsilon$, and conversely $q^* \gg p^* \iff \epsilon \gg J$: the inter-trade time $J^{-1}$ is much smaller (resp. larger) than the correlation time of the predictor $\epsilon^{-1}$ when the no-trade region is small (resp. large) compared to the typical scale of the predictor $p^*$, corresponding to small (resp. large) linear costs. 

Recall at this point that all these results are in fact $i$-dependent and that the parameter $\bar{J}$ describing the dynamics of the mean-field $R_t$ is obtained as a weighted average of the $J_i$ with weights $\beta_i^2 M_i^2$. We are now in position to derive how the no-trade thresholds $q^*_i$ are modified in the presence of a (small) risk term.

\section{Optimal thresholds for a double Ornstein-Uhlenbeck predictor} \label{sec:optimal_thresholds}

\subsection{The general setting}

We are thus faced with an effective single asset problem with a modified predictor that is the sum of two independent Ornstein-Uhlenbeck processes, one coming from the signal, and the other coming from the risk. This new predictor can be constructed from a two-dimensional stochastic process $X_{t}=\left(p_t,R_t\right)$, with $p_{t}$ and $R_{t}$ independent (due to the mean-field approximation). Its dynamics writes (in all the following sections, we again skip the index $i$):

\begin{equation}
    dX_{t}=\begin{bmatrix} 
-\epsilon & 0 \\
0 & -2\bar{J}
\end{bmatrix}X_{t}dt+\begin{bmatrix} 
\psi & 0 \\
0 & 2 \Sigma \sqrt{\bar{J}}
\end{bmatrix}dW_{t}=-AX_{t}dt+BdW_{t},
\end{equation}
where $W_{t}\in\mathbb{R}^2$ is a Brownian motion with independent components, 
$A=\begin{bmatrix} 
\epsilon & 0 \\
0 & 2\bar{J}
\end{bmatrix}$ and $B=\begin{bmatrix} 
\psi & 0 \\
0 & 2 \Sigma \sqrt{\bar{J}}
\end{bmatrix}$.

For our problem, only the combination $s_t=p_t - \theta R_t$ appears in the optimization problem, Eq. \eqref{opt_prob}. 
Adapting the arguments of \cite{Lataillade}, we expect that the optimal trading strategy is again of the ``bang-bang'' type between $+M$ and $-M$, but with thresholds on $p$ that become functions of the current value of $R_t$,  i.e.:
\begin{itemize}
    \item Switch to the maximum allowed position $+M$ when $p_t \geq q_+(R_t)$.
    \item Switch to the minimum allowed position $-M$ when $p_t \leq q_-(R_t)$.
    \item Otherwise, keep the previous position.
\end{itemize}
Following the steps of \cite{Lataillade}, the equilibrium between expected risk-adjusted gains and expected transaction costs can be expressed through some functions $\mathcal{P}_\pm(p,r)$ and $\mathcal{L}(p,r)$ defined as constrained path integrals: 
\begin{equation}
    \mathcal{L}(p,r)=	\int_{X_{0}=(p,r),\ -q_{-}(r)<p(t)<q_{+}(r),\forall t\in]0,T_1[}^{[p(T_1)\geq q_{+}(r)] \vee [p(T_1)\leq q_{-}(r)] }\left[\int_{0}^{T_1}w^{\top} X_{t} {\rm d}t\right]P(X|X_0)\mathcal{D}X,
\end{equation}
with $w^{\top} = (1,-\theta)$;
\begin{equation}
    \mathcal{P}_+(p,r)=	\int_{X_{0}=(p,r),\ -q_{-}(r)<p(t)<q_{+}(r),\forall t\in]0,T_1[}^{p(T_1)\geq q_{+}(r)} P(X|X_0)\mathcal{D}X;
\end{equation}
and
\begin{equation}
    \mathcal{P}_-(p,r)=	\int_{X_{0}=(p,r),\ -q_{-}(r)<p(t)<q_{+}(r),\forall t\in]0,T_1[}^{p(T_1)\leq q_{-}(r)} P(X|X_0)\mathcal{D}X,
\end{equation}
where $T_1$ is the first hitting time of one of the boundaries $q_\pm(r)$. Note that, by construction, $\mathcal{P}_-(p,r)+\mathcal{P}_+(p,r)=1$. Using these definitions, the conditions fixing the optimal no-trade boundaries are (see \cite{Lataillade}):
\begin{equation}
    \begin{cases}
        \mathcal{L}(p,r) = 2 \Gamma \mathcal{P}_-(p,r) & \text{when } p \rightarrow q_{+}(r) \\
        \mathcal{L}(p,r) = -2 \Gamma \mathcal{P}_+(p,r) & \text{when } p \rightarrow q_{-}(r).
    \end{cases}
\end{equation}
Equivalently, using the two-dimensional Brownian dynamics of $X_t$, these functions are the solutions of the following backward Kolmogorov equations:
\begin{equation}
    \label{vectorode} 
    \begin{cases}
        -\epsilon p\frac{\partial \mathcal{L}}{\partial p}-2\bar{J}r\frac{\partial \mathcal{L}}{\partial r}+\frac{\psi^{2}}{2}\frac{\partial^{2}\mathcal{L}}{\partial p^{2}}+2\Sigma^{2}\bar{J}\frac{\partial^{2}\mathcal{L}}{\partial r^{2}}=-p+\theta r\\
        -\epsilon p\frac{\partial \mathcal{P}_\pm}{\partial p}-2\bar{J}r\frac{\partial \mathcal{P}_\pm}{\partial r}+\frac{\psi^{2}}{2}\frac{\partial^{2}\mathcal{P}_\pm}{\partial p^{2}}+2\Sigma^{2}\bar{J}\frac{\partial^{2}\mathcal{P}_\pm}{\partial r^{2}}=0\\
    \end{cases}	
\end{equation}
with boundary conditions:
\begin{equation}
    \label{vectorode2} 
    \begin{cases}
        \mathcal{L}(q_{\pm}(r),r)=0\\
        \mathcal{P}_-(q_{-}(r),r)=\mathcal{P}_+(q_{+}(r),r)=1\\
        \mathcal{P}_-(q_{+}(r),r)=\mathcal{P}_+(q_{-}(r),r)=0.
    \end{cases}	
\end{equation}
These equations are hard to solve in full generality. In the following, we obtain the exact solution in the special symmetry point $\epsilon=2\bar{J}$ (when the mean-field $R_t$ and the predictor $p_t$ have the same auto-correlation time), and an approximate solution in the small $\theta$ limit, where the risk contribution is a small correction to predictors.

\subsection{The symmetric case: an exact solution}

Let us study the special case $\epsilon=2\bar{J}$ for which an exact solution is available. In this case, the modified predictor $s_t=p_{t}-\theta R_{t}$ is a one-dimensional Ornstein-Uhlenbeck process, namely: 
\begin{equation}
    ds_{t}=-\epsilon s_t dt +\tilde{\psi}dW_{t},
\end{equation}
with $\tilde{\psi}:=\sqrt{\psi^{2}+2\epsilon\Sigma^2\theta^{2}}$. Hence we are back to the problem considered in \cite{Lataillade}. In the interesting continuous time regime $\Gamma \epsilon^{3/2} 
< \tilde{\psi} < \Gamma$, we therefore obtain a threshold for $s_t$:
\begin{equation}\label{eq_q}
    \tilde q^* = \left(\frac{3}{2}\Gamma \tilde{\psi}^2 \right)^{1/3}.
\end{equation}
or, translated into thresholds for $p_t$:
\begin{equation}\label{eq_q_cases}
    \begin{cases}
    q_+(r) = \tilde q^* + \theta r \\
    q_-(r) = -\tilde q^* + \theta r.
    \end{cases}
\end{equation}
These expressions have a simple interpretation: in the case $\lambda > 0$, when the net position is positive ($R_t > 0$), any positive signal to buy must exceed an increased threshold $\tilde q^* \to \tilde q^* + \theta R_t$ for taking some extra risk, and vice-versa when the trade is such that $|R_t|$ is reduced.

\subsection{General case: approximate solution} \label{thm:general}

In some regimes, we can find approximate solutions of Eq. \eqref{vectorode}. In particular, we are interested in the case in which $\theta \to 0$, that is to say when trading is mostly driven by the predictors and not by the risk constraint. In this case, we also know from Section \ref{sec:th_rates} how to calculate $\bar{J}$ exactly. We will assume again that we are in the intermediate, continuous-time regime $\Gamma \epsilon^{3/2} < {\psi} < \Gamma$ for all assets (see Eq. \eqref{lataillade_threshold}). We introduce $q^* = \left( \frac{3}{2} \Gamma \psi^2 \right)^\frac{1}{3}$ as the threshold without the risk constraint, which is such that $q^* \ll p^*:=\psi/\sqrt{\epsilon}$ and hence, from Eq. \eqref{J_approx}, $\epsilon \ll \bar{J}$. In this regime, the predictor beats its costs with some significant probability (see \cite{Lataillade}), but is much slower than the risk component. 

We are now ready to state our main result. Assume that $\theta \Sigma \ll p^*$ and $\epsilon \ll \bar{J}$. Then, neglecting terms of order $(q^*/p^*)^4$, 
the optimal thresholds are given by: 
\begin{equation}
    \begin{cases}
    q_+(r) \approx q_1 + S \theta r \\
    q_-(r) \approx -q_1 + S \theta r,
    \end{cases}
\end{equation}
with:
\begin{equation}\label{final_result}
q_1:=q^*\left(1-\frac{1}{3}\frac{q^{*2}}{p^{*2}}\right); \qquad
    S := \left[\left(1-\frac{q^{*2}}{p^{*2}}\right)\left(1+\frac{2\bar{J}}{\epsilon}\frac{q^{*2}}{p^{*2}}\right)\right]^{-1}.
\end{equation}

See Appendix \ref{app:general_proof} for a proof. Observe that when $2\bar{J}=\epsilon$, we find that $S=1+O(a^2q^{*4})$, and we recover precisely the result for the symmetric case derived in the previous section.

More generally, the slope $S$ of the risk correction term $S\theta R_t$ decreases towards $0$ as $1/\bar{J}$. As $\bar{J}$ increases, the memory of the mean-field process shortens. In the limit where the risk contribution to the modified predictor becomes a white noise process, it must indeed become irrelevant in determining the optimal trading strategy ($S \to 0$). 

\section{Fixing the risk-aversion parameter}
\label{sec:fixing_lambda}
We now come back to the issue of the realized risk in our mean-field formalism. We recall that, in the stationary limit, the expected risk per asset  $\mathcal{R}$ is: 
\begin{equation}
    \mathcal{R} = \frac{1}{N}\mathbb{E}[\vec \pi^\top \mathbf{C} \vec \pi] =  \frac{1}{N} \mathbb{E}\left[\sum_{i=1}^{N} C_{ii} (\pi^{i})^2 + \sqrt{N} \sum_{i=1}^{N} \beta_i \pi^{i} R\right].
\end{equation}
Now, since the no-trade region threshold is shifted by an amount $S\theta R$, there appears some correlations between positions $\pi^i$ and the mean-field $R$. Using
\begin{equation}
    \mathbb{E}[\pi^i R]=M_i \mathbb{E}\left[R \left(\mathbb{P}[\pi^i=M_i|R] - \mathbb{P}[\pi^i=-M_i|R]\right)\right],
\end{equation}
and expanding to first order in $\theta$, one finds:
\begin{equation}
    \mathbb{E}[\pi^i R] \approx - 2 M_i S_i \theta_i \rho_i \mathbb{E}[R^2],
\end{equation}
where $\rho_i$ is the probability density of the predictor at its unperturbed threshold $q^*_i$. Now, in order to have a result valid in the whole region $\lambda=O(1)$, we have to estimate $\mathbb{E}[R^2]$ self-consistently. 
Writing
\begin{equation}
    \mathbb{E}[R^2] = \mathbb{E}\left[\frac{1}{N} \sum_i \beta_i^2 M_i^2 + 
    \frac{1}{\sqrt{N}} \sum_i \beta_i \pi^i R\right],
\end{equation}
one gets:
\begin{equation}
    \mathbb{E}[R^2] = \Sigma^2 - \lambda \Xi^2 \mathbb{E}[R^2] \implies
    \mathbb{E}[R^2] = \frac{\Sigma^2}{1+ \lambda \Xi^2},
\end{equation}
where $\Sigma^2 := \sum_i M_i^2 \beta_i^2/N$ and $\Xi^2:=2\sum_i M_i \beta_i^2 S_i \rho_i/N$ are intensive quantities. 

Hence the risk per asset is given by
\begin{equation}\label{eq_final_risk}
    \mathcal{R} = \mathcal{R}_0 - \frac{{\lambda}\Sigma^2 \Xi^2}{1+ \lambda \Xi^2},\qquad \mathcal{R}_0:=\frac{1}{N} \sum_i C_{ii} M_i^2,
\end{equation}
where $\mathcal{R}_0$ is the expected risk per asset in the fully decoupled problem, $\lambda=0$.

Eq. \eqref{eq_final_risk} shows that the scaling $\lambda=O(1)$ indeed allows us to tune the realized risk. Note that one has to choose a negative risk aversion parameter in order to realize a risk larger than $\mathcal{R}_0$. 

Note that, writing $\mathcal{R}^{\text{min}}=\sum_{i} \sigma_i^2 M_i^2/N$ the minimum achievable risk (in the case $\beta_i=0 \ \forall i$), one gets:
\begin{equation}
    \mathcal{R} = \mathcal{R}^{\text{min}} + \frac{\Sigma^2}{1+\lambda \Xi^2},
\end{equation}
which confirms that, $\mathcal{R}\rightarrow \mathcal{R}^{\text{min}}$ when the risk-aversion $\lambda$ goes to $+\infty$ (which is a valid regime as long as $\lambda=o(\sqrt{N})$).

\section{Numerical Simulations}

\subsection{Dynamics of the mean-field}

\begin{figure}[htbp]
\centering
\includegraphics[width=0.53\textwidth]{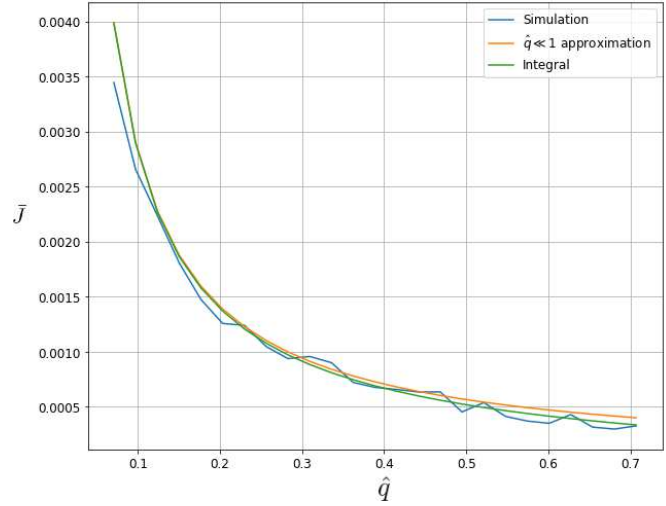}
\caption{Simulated $\bar{J}$ (blue), approximation (orange) and exact formula for $\bar{J}$ (green).}
\label{fig:J_simulation}
\end{figure}

The results of Section \ref{sec:mean_field_dynamics} and Section \ref{sec:optimal_thresholds} are easily confirmed with numerical simulations in the case where the predictor often beats its threshold ($q^*\ll p^*$). We simulate the case where $N=100$, predictors have identical dynamics (with $\psi=\epsilon=10^{-3}$), identical trading costs $\Gamma$, and an effective maximum position $\beta_i M_i$ common to all stocks, set to unity. We then run the ``bang-bang'' strategy on these predictors with a given threshold, and compute the dynamics of the average position by maximum likelihood estimation. Results are shown in Fig. \ref{fig:J_simulation}. We compare three ways to recover $\bar{J}$: via simulation of $R_t$, via the approximation valid when $\hat{q}\ll 1$ (Eq. \eqref{J_approx}), or via a direct computation using the integral formulation Eq. \eqref{FullJ}.

\subsection{Dependence of the threshold on risk}

\begin{figure}[htbp]
\centering
\includegraphics[width=0.8\textwidth]{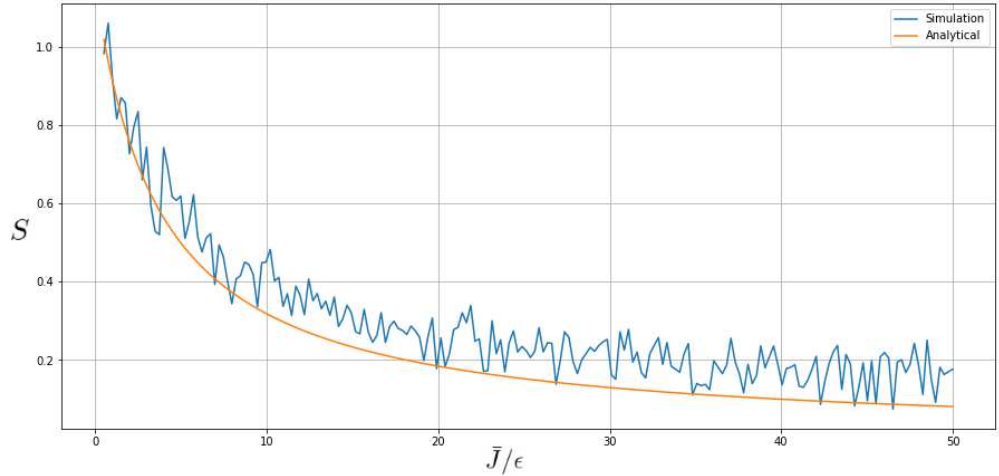}
\caption{Simulated (blue) vs. analytical (orange) optimal $S$ as a function of $\bar{J}/\epsilon$}
\label{fig:optimalE}
\end{figure}

To check the correctness of our main analytical result, Eq. \eqref{final_result}, we ran simulations to determine the optimal correction $S$ to the original threshold $q^*$ 
when the risk correction is present, for various values of $\bar{J}$. Results are shown in Fig. \ref{fig:optimalE}. Our set of parameters is:
\begin{equation}
    \beta M = 1;
    \qquad
    \psi = 10^{-3};
    \qquad
    \epsilon = 10^{-3};
    \qquad
    \Gamma = 1;
    \qquad
    \theta = 5 \times 10^{-3}.
\end{equation}
With this set of parameters, we have:
\begin{equation}
    (q^*/p^*)^2 \sim 0.1 \ll 1;
    \qquad
    (\theta M/p^*)^2 \sim 0.025 \ll 1;
    \qquad
    \psi/\Gamma\epsilon^{3/2} \sim 30 \gg 1.
\end{equation}
The first two regimes match our assumptions. The third regime is needed (see \cite{Lataillade}) to get $q^* \approx \left(\frac{3}{2} \Gamma \psi^2 \right)^\frac{1}{3}$.
The algorithm to find the optimal correction $S$ for a fixed value of $\bar{J}$ is similar to the one used in \cite{Lataillade}:

\begin{itemize}
    \item We choose a set of values $S_1,...,S_k$ distributed over a range that contains the theoretical optimal $S$.
    \item We generate two long random paths $(p_t)_{t \in [0,T]}$ with the dynamics of Eq. \eqref{p_dynamics} and $(R_t)_{t \in [0,T]}$ with the dynamics of Eq. \eqref{R_dynamics}. We choose $T=2,500,000$.
    \item For each value of $S_n$, we simulate the behaviour of the corresponding strategy (using $q_\pm(r) = \pm q_1+S\theta r$), and obtain the P\&L, taking into account the risk-corrected gains $(p_t-\theta R_t)\pi_t$ and the costs $\Gamma |\pi_{t}-\pi_{t-1}|$.
    \item We select the value $S_j$ with the maximum total P\&L, and choose new values $S^\prime_1,...,S^\prime_n$ for the possible optimal correction, uniformly distributed around $S_j$, and restart the same process.
\end{itemize}

We repeat this loop three times in Fig. \ref{fig:optimalE}. Remark that the agreement is not perfect, because the regime $q_\pm(r) \ll p^*$ is not totally fulfilled ($q_\pm(r) \sim 0.3 p^*$ here). Moreover, when $\bar{J}$ tends towards $\epsilon/2$, we saw that $S=1+O((q^*/p^*)^4)$, and so the correction does not matter much. Hence, as seen from the left part of the graph, the simulation becomes closer to the analytical result.

\section{Conclusion}
In this paper, we developed an extension to \cite{Lataillade} in a multi-asset setting with a quadratic risk constraint. The coupling between the assets induced by the risk constraint makes this problem difficult to solve in the general case. However, in the limit of a large number of assets, a mean-field approach allowed us to rewrite the problem as $N$ single-asset problems, where each individual predictor is a linear combination of the original predictor and the instantaneous global position on all assets. The dynamics of the global position is shown to be Ornstein-Uhlenbeck. When the risk control is small, we showed that the optimal strategy is similar to the one of the single-asset case, except that the thresholds become dependent on the instantaneous global position. We described this effect both quantitatively and qualitatively through numerical simulations. In this regime with moving thresholds, we are also able to link the risk aversion parameter to the target risk, which allows to tune the realized risk easily.

This result could be extended in many ways, for instance by considering the case of more general correlation matrices. We believe that an extension of our $1$-factor risk model to a $k$-factor risk model (corresponding to block-hierarchical covariance matrices) should still be analytically solvable using the same ideas, with $k$ independent mean-field risk terms each with its own Ornstein-Uhlenbeck dynamics. Another direction would be to move beyond first order in the risk aversion parameter $\theta$. This would first require obtaining the dynamics of the mean-field beyond zero-th order. It would be interesting to prove mathematically that our mean-field approximation is exact as $N \to \infty$, and that the optimal strategy remains of the bang-bang type to all orders in $\theta$. Finally, we believe that our mean-field approach naturally extends to other, more general forms of costs for which we can derive the optimal solution for a single-asset problem (e.g. in the presence of linear costs and small quadratic costs \cite{Rej}). The dynamics of the mean-field might be different, but the approach we use to disentangle the multi-asset formulation should remain valid.

\vskip 0.5cm
\section*{Acknowledgments}

We thank F. Altarelli, J. de Lataillade, A. Maillard and J. Muhle-Karbe for interesting discussions on this topic. We specially thank J. Muhle-Karbe for carefully reading our manuscript and suggesting many improvements, and an unknown referee for insisting that we clarify the scaling of the risk aversion parameter.

\bibliographystyle{alpha}
\newcommand{\etalchar}[1]{$^{#1}$}

\newpage

\appendix
\addcontentsline{toc}{section}{Appendices}
\section*{Appendices}

\section{Conditional value of \texorpdfstring{$\pi^i$}{pi} knowing \texorpdfstring{$R$}{R}} \label{app:cond_pi_R}

By definition, the mean-field $R$ is given by:
\begin{equation}
    R = \frac{1}{\sqrt{N}} \sum_{i=1}^N \beta_i \pi^i = \frac{1}{\sqrt{N}} \sum_{i=1}^N S_i \beta_i M_i,
\end{equation}
where $S_i = \pm 1$ with probability $1/2,1/2$. From the Central Limit Theorem, the unconditional distribution of $R$ is therefore, at large $N$, Gaussian with
zero mean and variance $\Sigma^2:=\sum_i \beta_i^2 M_i^2/N$.

Conditionally to a certain $S_1$, the distribution of $R$ is therefore
\begin{equation}
    P(R|S_1) = \frac{1}{\sqrt{2 \pi \Sigma^2}} e^{-\frac{\left(R - S_1 \beta_1 M_1/\sqrt{N}\right)^2}{2 \Sigma^2}}
    \underset{N \to \infty}{\approx} \frac{1}{\sqrt{2 \pi \Sigma^2}} e^{-\frac{R^2}{2 \Sigma^2}} \left(1 + \frac{R S_1 \beta_1 M_1}{\Sigma^2 \sqrt{N}}\right).
\end{equation}

Now, using Bayes, one has
\begin{equation}
    P(S_1|R) = \frac12 \frac{P(R|S_1)}{P(R)} = \frac12 \left(1 + \frac{R S_1 \beta_1 M_1}{\Sigma^2 \sqrt{N}}\right).
\end{equation}
Since ``1'' here plays no special role, the same result holds for any $i$, as announced in the main text.

\section{Asymptotic expansion for \texorpdfstring{$J$}{J}} \label{app:J_expansion}

We expand the equation for $J$ in two different regimes, namely $\hat{q}\ll 1$ and $\hat{q}\gg 1$. Using $\hat{q}=q^*/p^*$ with $p^{*2}=1/a$ leads to:
\begin{equation}
    \frac{1}{J}=\frac{2\xi^{2}}{\psi^{2}}\left[\int_{-\hat{q}}^{\hat{q}}e^{-p^{2}}\left(\int_{-\hat{q}}^{p}e^{x^{2}}dx\right)dp+\int_{\hat{q}}^{+\infty}e^{-p^{2}}dp\int_{-\hat{q}}^{\hat{q}}e^{p^{2}}dp\right] = \frac{2\xi^{2}}{\psi^{2}}\left[I_{1}+I_{2}\right].
\end{equation}
In the case $\hat{q}\ll 1$, we easily find:
\begin{equation}
    \begin{cases}
    I_{1}=2\hat{q}^{2}+O(\hat{q}^{6})\\
    I_{2}=\sqrt{\pi}\hat{q}-2\hat{q}^{2}+O(\hat{q}^{3}).
    \end{cases}	
\end{equation}
Hence:
\begin{equation}
    \frac{1}{J}=2\sqrt{\pi}\frac{\hat{q}}{\epsilon}.
\end{equation}
In the case  $\hat{q}\gg 1$, we get, by integration by parts (where $C$ is a constant):
\begin{align}
    I_{2} & = 2\int_{\hat{q}}^{+\infty}e^{-p^{2}}dp\int_{0}^{\hat{q}}e^{p^{2}}dp \\
    & = 2\left[\frac{e^{-\hat{q}^{2}}}{2\hat{q}}-\int_{\hat{q}}^{+\infty}\frac{e^{-p^{2}}}{2p^{2}}dp\right]\left[C+\frac{e^{\hat{q}^{2}}}{2\hat{q}}+\int_{1}^{\hat{q}}\frac{e^{p^{2}}}{2p^{2}}dp\right] \\
    & = 2\left[\frac{e^{-\hat{q}^{2}}}{2\hat{q}}+O\left(\frac{e^{-\hat{q}^{2}}}{\hat{q}^{3}}\right)\right]\left[\frac{e^{\hat{q}^{2}}}{2\hat{q}}+O\left(\frac{e^{\hat{q}^{2}}}{\hat{q}^{3}}\right)\right] \\
    & = \frac{1}{2\hat{q}^{2}}+O\left(\frac{1}{\hat{q}^{4}}\right).
\end{align}
We can now expand $I_{1}$ by cutting the interior integral at $x=0$:
\begin{align}
    I_{1} & = \int_{-\hat{q}}^{\hat{q}}e^{-p^{2}}dp\int_{-\hat{q}}^{0}e^{x^{2}}dx + \int_{-\hat{q}}^{\hat{q}}e^{-p^{2}}\left(\int_{0}^{p}e^{x^{2}}dx\right)dp \\
    & = \left[\sqrt{\pi}+O\left(\frac{e^{-\hat{q}^{2}}}{\hat{q}}\right)\right]\left[\frac{e^{\hat{q}^{2}}}{2\hat{q}}+O\left(\frac{e^{\hat{q}^{2}}}{\hat{q}^{3}}\right)\right]+2\int_{0}^{\hat{q}}e^{-p^{2}}\left(\int_{0}^{p}e^{x^{2}}dx\right)dp \\
    & = \frac{\sqrt{\pi}e^{\hat{q}^{2}}}{2\hat{q}}+O\left(\frac{e^{\hat{q}^{2}}}{\hat{q}^{2}}\right).
\end{align}
since $\int_{0}^{\hat{q}}e^{-p^{2}}\left(\int_{0}^{p}e^{x^{2}}dx\right)dp\leq\frac{\hat{q}^{2}}{2}=O\left(\frac{e^{\hat{q}^{2}}}{\hat{q}^{3}}\right)$.

Hence:
\begin{equation}
    \frac{1}{J}=\sqrt{\pi}\frac{e^{\hat{q}^{2}}}{\epsilon\hat{q}}.
\end{equation}

\section{Proof of main result in \ref{thm:general}}\label{app:general_proof}

Justifying this result requires to also be in the regime $aq_\pm(r)^2 \ll 1$. However, remember that we proved earlier that the regimes $\epsilon \ll \bar{J}$ and $aq_\pm(r)^2 \ll 1$ were equivalent. Using the symmetry properties of $\mathcal{L}$ and $\mathcal{P}_-$ under $r \to -r$ and $p \to -p$, we can expand these functions as: 
\begin{equation}
    \begin{cases}
        \mathcal{L}(p,r) = A \theta r + B p + C \theta r p^2 + D p^3 \\
        \mathcal{P}_-(p,r) = A^{\prime\prime} + A^\prime \theta r + B^\prime p + C^\prime \theta r p^2 + D^\prime p^3.
    \end{cases}
\end{equation}
Even though we defined the function $\mathcal{P}_+$, we do not need to expand it, since $\mathcal{P}_- = 1-\mathcal{P}_+$. We can plug these expressions in the PDEs, and look at the first order terms in $p$ and in $\theta r$ (we expanded the two functions at a sufficient order in $p$ and $\theta r$ so that we catch all the terms at first order in $p$ and $\theta r$ in the PDE). We obtain:
\begin{equation}
    \begin{cases}
        p\left(-\epsilon B + 3 D \psi^2\right) + \theta r \left( C\psi^2 - 2\bar{J}A\right) = -p+\theta r \\
        p\left(-\epsilon B^\prime + 3 D^\prime \psi^2\right) + \theta r \left( C^\prime\psi^2 - 2\bar{J}A^\prime \right) = 0.
    \end{cases}
\end{equation}
Subsequently, we get four conditions on the constants:
\begin{equation} \label{dvpt_thr_constants}
    \begin{cases}
        \epsilon B - 3 D \psi^2 = 1 \\
        C\psi^2 - 2\bar{J}A = 1 \\
        \epsilon B^\prime - 3 D^\prime \psi^2 = 0 \\
        C^\prime\psi^2 - 2\bar{J}A^\prime = 0.
    \end{cases}
\end{equation}
Using the boundary conditions gives:
\begin{equation} \label{dvpt_thr_boundary}
    \begin{cases}
        A \theta r + B q_{\pm}(r) + C \theta r q_{\pm}(r)^2 + D q_{\pm}(r)^3 = 0 \\
        A^{\prime\prime} + A^\prime \theta r + B^\prime q_{-}(r) + C^\prime \theta r q_{-}(r)^2 + D^\prime q_{-}(r)^3 = 1 \\
        A^{\prime\prime} + A^\prime \theta r + B^\prime q_{+}(r) + C^\prime \theta r q_{+}(r)^2 + D^\prime q_{+}(r)^3 = 0.
    \end{cases}
\end{equation}
Simultaneously, we use the fact that:
\begin{equation}
    \frac{\frac{\partial L}{\partial p}(p,r)}{\frac{\partial \mathcal{P}_-}{\partial p}(p,r)} \xrightarrow[p\rightarrow q_{\pm}(r)]{} 2\Gamma.
\end{equation}
This re-writes:
\begin{equation} \label{dvpt_thr_partial}
    \begin{cases}
        B + 2C \theta r q_{-}(r) + 3 D q_{-}(r)^2 = 2\Gamma \left( B^\prime + 2 C^\prime \theta r q_{-}(r) + 3 D^\prime q_{-}(r)^2\right) \\
        B + 2C \theta r q_{+}(r) + 3 D q_{+}(r)^2 = 2\Gamma \left( B^\prime + 2 C^\prime \theta r q_{+}(r) + 3 D^\prime q_{+}(r)^2\right).
    \end{cases}
\end{equation}
At first order in $\theta r$, let us define:
\begin{equation}
    q_{\pm}(r) = q_{\pm,0} + S_{\pm} \theta r.
\end{equation}
Expanding the boundary conditions Eq. \eqref{dvpt_thr_boundary} to first order in $\theta r$ yields:
\begin{equation} \label{dvpt_thr_boundary_dvp}
    \begin{cases}
        B + D q_{\pm,0}^2 = 0 \\
        A + B S_{\pm} + C q_{\pm,0}^2 + 3 D S_{\pm} q_{\pm,0}^2 = 0 \\
        A^{\prime\prime} + B^\prime q_{-,0} + D^\prime q_{-,0}^3 = 1 \\
        A^{\prime\prime} + B^\prime q_{+,0} + D^\prime q_{+,0}^3 = 0 \\
        A^\prime + B^\prime S_{\pm} + C^\prime q_{\pm,0}^2 + 3 D^\prime S_{\pm} q_{\pm,0}^2 = 0.
    \end{cases}
\end{equation}
At first order in $\theta r$, Eq. \eqref{dvpt_thr_partial} re-writes:
\begin{equation} 
    \begin{cases}
        B + 2C \theta r q_{-,0} + 3 D \left(q_{-,0}^2+2q_{-,0}S_- \theta r\right) = 2\Gamma \left[ B^\prime + 2 C^\prime \theta r q_{-,0} + 3 D^\prime \left(q_{-,0}^2+2q_{-,0}S_- \theta r\right)\right] \\
        B + 2C \theta r q_{+,0} + 3 D \left(q_{+,0}^2+2q_{+,0}S_+ \theta r\right) = 2\Gamma \left[ B^\prime + 2 C^\prime \theta r q_{+,0} + 3 D^\prime \left(q_{+,0}^2+2q_{+,0}S_+ \theta r\right)\right],
    \end{cases}
\end{equation}
or in a more simplified version (separating the order $0$ and order $1$ in $\theta r$):
\begin{equation} \label{dvpt_thr_partial_dvp}
    \begin{cases}
        B + 3 D q_{\pm,0}^2 = 2\Gamma \left( B^\prime + 3 D^\prime q_{\pm,0}^2 \right) \\
        C + 3 D S_\pm = 2\Gamma \left[C^\prime + 3 D^\prime S_\pm \right].
    \end{cases}
\end{equation}
The first boundary condition in Eq. \eqref{dvpt_thr_boundary_dvp} coupled to the first equation in Eq. \eqref{dvpt_thr_constants} gives:
\begin{equation}
    \begin{cases}
        B = \frac{q_{\pm,0}^2}{3\psi^2} \left(1 - \frac{1}{3}a q_{\pm,0}^2 \right) \\
        D = \frac{1}{3\psi^2} \left(-1 + \frac{1}{3}a q_{\pm,0}^2 \right).
    \end{cases}
\end{equation}
At order $0$ in $aq_{\pm,0}^2$, we immediately obtain that $q_{-,0} = - q_{+,0}$. In the following, we will denote, as in the main text, $q^* = q_{+,0} = - q_{-,0}$. 
Using the third and fourth boundary conditions in Eq. \eqref{dvpt_thr_boundary_dvp} as well as the third equation in Eq. \eqref{dvpt_thr_constants}, we get:
\begin{equation}
    \begin{cases}
        B^\prime = \frac{1}{2q^*} \left(-1 + \frac{1}{3}a q^{*2} \right) \\
        D^\prime = \frac{a}{6q^*} \left(-1 + \frac{1}{3}a q^{*2} \right).
    \end{cases}
\end{equation}
Now we can plug these expressions in the first equation in Eq. \eqref{dvpt_thr_partial_dvp} to get, at order $0$ in $aq^{*2}$:
\begin{equation}
    \frac{q^{*2}}{3\psi^2} - \frac{q^{*2}}{\psi^2} = 2\Gamma \frac{1}{2q^*}.
\end{equation}
It follows that $q^* = \left(\frac{3}{2} \Gamma \psi^2 \right)^\frac{1}{3}$, as expected. The expressions of $B^\prime, D^\prime$ simplify further to:
\begin{equation}
    \begin{cases}
        B^\prime = \frac{q^{*2}}{3\Gamma\psi^2} \left(-1 + \frac{1}{3}a q^{*2} \right) \\
        D^\prime = \frac{aq^{*2}}{9\Gamma\psi^2} \left(-1 + \frac{1}{3}a q^{*2} \right).
    \end{cases}
\end{equation}
Let us now obtain a more precise expression of $q^*$ (at first order in $aq^{*2}$). We will write $q_1=q^* \left(1+\lambda aq^{*2}\right)$, the ``new'' $q^*$ that we want to obtain for the following computation. The first equation of Eq. \eqref{dvpt_thr_partial_dvp} gives:
\begin{equation}
    \frac{q_1^2}{3\psi^2} \left[1-\frac{1}{3}aq_1^2 -3 +aq_1^2\right] = 2\Gamma\left[-\frac{1}{2q_1}\left(1-\frac{1}{3}aq_1^2\right)+\frac{1}{2}aq_1\left(1-\frac{1}{3}aq_1^2\right)\right].
\end{equation}
This quickly yields:
\begin{equation}
    q_1^3 = q^{*3} \left(1-aq^{*2}\right).
\end{equation}
It easily follows that $\lambda = -\frac{1}{3}$. For now, let us continue with the order $0$ condition.

We can re-write the second and fifth boundary conditions in Eq. \eqref{dvpt_thr_boundary_dvp} using the formulas for $B,B^\prime,D,D^\prime$ as well as the second and fourth equations in Eq. \eqref{dvpt_thr_constants}. We will denote $b=\frac{\bar{J}}{\psi^2}=a\frac{\bar{J}}{\epsilon}$. We also quickly obtain that $S_\pm = S_+ = S_-$ and we will denote $S_\pm = S$:
\begin{equation}
    \begin{cases}
        \frac{C\psi^2-1}{2\bar{J}} - \frac{2S q^{*2}}{3\psi^2} \left(1-\frac{1}{3}a q^{*2}\right) + Cq^{*2} = 0 \\
        \frac{C^\prime \psi^2}{2\bar{J}} - \frac{Sq^{*2}}{3\Gamma\psi^2} \left(1+\frac{2}{3}a q^{*2}\right) + C^\prime q^{*2} = 0,
    \end{cases}
\end{equation}
which yields the values of $C, C^\prime$:
\begin{equation}
    \begin{cases}
        C = \left[\frac{b}{\bar{J}} + \frac{4Sbq^{*2}}{3\psi^2} \left(1-\frac{1}{3}aq^{*2}\right)\right]\frac{1}{1+2bq^{*2}} \\
        C^\prime = \frac{2S bq^{*2}}{3\Gamma \psi^2} \frac{1+\frac{2}{3}a q^{*2}}{1+2bq^{*2}}.
    \end{cases}
\end{equation}
We can finally plug that into the second equation in Eq. \eqref{dvpt_thr_partial_dvp}:
\begin{multline}
    \frac{b}{\bar{J}} + \frac{4Sbq^{*2}}{3\psi^2}\left(1-\frac{1}{3}aq^{*2}\right) + \frac{S}{\psi^2} \left(-1+\frac{1}{3}aq^{*2}\right)\left(1+2bq^{*2}\right) \\ =\frac{4Sbq^{*2}}{3\psi^2}\left(1+\frac{2}{3}aq^{*2}\right)+\frac{2Saq^{*2}}{3\psi^2}\left(-1+\frac{1}{3}aq^{*2}\right)\left(1+2bq^{*2}\right).
\end{multline}
Multiplying by $\psi^2$ and grouping the terms with $S$ yields:
\begin{multline}
     1 = S \left[ -\frac{4}{3}bq^{*2} \left(1-\frac{1}{3}aq^{*2}\right) - \left(-1+\frac{1}{3}aq^{*2}\right)\left(1+2bq^{*2}\right) \right. \\ \left. +\frac{4}{3}bq^{*2}\left(1+\frac{2}{3}aq^{*2}\right)+\frac{2}{3}aq^{*2}\left(-1+\frac{1}{3}aq^{*2}\right)\left(1+2bq^{*2}\right) \vphantom{\int_1^2} \right].
\end{multline}
In the case where $\bar{J}\gg \epsilon$, we can neglect the terms of the type $a^2q^{*4}$ compared to those of the form $bq^{*2} \cdot aq^{*2}$ (if we had to consider the terms of the form $a^2q^{*4}$, other terms of this order that we neglected earlier in the development might appear, and we would also have to develop the functions $\mathcal{L}$ and $\mathcal{P}_-$ at a higher order). In this regime, we obtain a simple formula for $S$: 
\begin{equation}
     S = \frac{1}{1-aq^{*2}+2bq^{*2}\left(1-\frac{1}{3}aq^{*2}\right)}.
\end{equation}
This formula should be amended to account for first order corrections to $q^*$, which transforms the $2bq^{*2}$ term into $2bq^{*2}(1-\frac{2}{3}aq^{*2})$. Hence we finally obtain:
\begin{equation}
     S = \frac{1}{1-aq^{*2}} \frac{1}{1+2bq^{*2}} = \frac{1}{1-aq^{*2}} \frac{1}{1+\frac{2\bar{J}}{\epsilon}aq^{*2}},
\end{equation}
which reproduces $S = 1$ to leading order when $b=a/2$.

\end{document}